\documentclass{article}
\usepackage[utf8]{inputenc}
\usepackage{biblatex}
\usepackage{hyperref}
\newcommand{\footremember}[2]{%
    \footnote{#2}
    \newcounter{#1}
    \setcounter{#1}{\value{footnote}}%
}
\newcommand{\footrecall}[1]{%
    \footnotemark[\value{#1}]%
}

\title{Vaccine Credential Technology Principles}
\author{Divya Siddarth\footremember{MS}{Microsoft} \and Vi Hart\footremember{MSR}{Microsoft Research} \and Bethan Cantrell\footrecall{MS} \and Kristina Yasuda\footrecall{MS} \and Josh Mandel\footrecall{MSR}\footremember{BCH}{Computational Health Informatics Program, Boston Children’s Hospital}\footremember{HMS}{Departments of Biomedical Informatics, Harvard Medical School} \and Karen Easterbrook\footrecall{MSR}}
\date{May 2021}

\begin{document}
\sffamily
\maketitle

\section{Executive Summary}

The historically rapid development of effective COVID-19 vaccines has policymakers facing evergreen public health questions regarding vaccination records and verification. Governments and institutions around the world are already taking action on digital vaccine certificates, including guidance and recommendations from the European Commission,\cite{EUguidance} the WHO,\cite{WHOguidance} and the Biden Administration. 

These could be encouraging efforts: an effective system for vaccine certificates could potentially be part of a safe return to work, travel, and daily life, and a secure technological implementation could improve on existing systems to prioritize privacy, streamline access, and build for necessary interoperability across countries and contexts. However, vaccine credentials are not without potential harms, and, particularly given major inequities in vaccine access and rollout, there are valid concerns that they may be used in ineffective or exclusionary ways that exacerbate inequality, allow for discrimination, violate privacy, and assume consent. While the present moment calls for urgency, we must also acknowledge that choices made in the vaccine credentialing rollout for COVID-19 are likely to have long-term implications, and must be made with care. 

In this paper, we outline potential implementation and ethical concerns that may arise from tech-enabled vaccine credentialing programs now and in the future, and discuss the technological tradeoffs implicated in these concerns. We suggest a set of principles that, if adopted, may mitigate these concerns, forestall preventable harms, and point the way forward; the paper is structured as a deep dive into each of these principles:

\begin{enumerate}
\item \textbf{Access and Equity.} We commit to working to understand and design for those who may not want or be able to participate in the vaccine credentialing process, including those who may not want to or be able to access vaccines.
\item \textbf{Choice and Consent.} We commit to maintaining active oversight to ensure that individuals and organizations retain freedom of choice and that no data is shared without consent.
\item \textbf{Technology-enabled, not technology-first.} We are committed to building technology scoped to the problem at hand, and working with affected communities and institutions to thoughtfully navigate their needs and realities.
\item \textbf{Privacy and security.} We commit to robust privacy and security guarantees, including transparent communication to credential holders. We will not sacrifice privacy for the sake of urgency.
\item \textbf{Interoperability.} We commit to enabling health data exchange that uses a non-proprietary, open, and accessible standard, and working with involved institutions to iterate on and maintain this standard, which is necessary for public health, private sector collaboration, and individual control over health data.
\item \textbf{Continued Transparency, Oversight, and Evolution.} We commit to continued learning, transparency, and communication as the situation evolves, paired with legal and ethical accountability mechanisms involving public health authorities and existing oversight bodies. 
\end{enumerate}

\section{Introduction}

Since initial discussions of technological interventions for COVID-19 in early 2020, we have had the opportunity to see technology such as exposure notifications implemented in the real world, leading to learnings about the role of technology in supporting an equitable, privacy-preserving, and successful pandemic response. One major outcome of these learnings has been a deeper understanding of the ways our technical interventions should be aligned with and in service of broader social outcomes, through partnering with existing institutions, communities, and organizations. A second major lesson is the need to ensure accessibility and transparency to the public to build trust, and determine when (and if) technological solutions are best suited for the task at hand. These insights and additional context should be applied in disentangling the recent surge of interest in COVID vaccine credentials. 

While the idea of vaccine credentials is not new, the present context introduces new and pressing concerns about potential negative effects of this technology being rushed into deployment in response to COVID. Proactive use of intentionally designed technology towards the purpose of vaccine credentialing could potentially result in a faster and safer return to work, travel, and other aspects of daily life, and thus is well worth investigating. However, any use of credentialing technology must protect individual rights and safety, and also not exacerbate the very real community- and economic-based divides that have already emerged throughout the pandemic.  

Thus, as we consider how and whether to develop this technology, it is crucial to do so in a manner aligned with a deep understanding of inherent tradeoffs and complexities. In this paper, we aim to share our thinking as researchers in this space, with a range of firsthand experience, both through involvement in early policy-setting regarding exposure notifications, as well as via years of navigating many of the complexities and concerns that emerge with proposed short-term technical interventions in large-scale societal problems. Our hope is that our experience may guide others’ explorations, particularly technologists; our aim is not to adjudicate whether these systems should be developed, but rather to understand possible ethical and implementation concerns, and point the way forward to a robust and effective ecosystem. For communicative clarity, the paper is structured as a deep dive into each of our proposed guiding principles.

\section{Verifiable Credentials for Vaccines}

A successful end-to-end credentialing system allows a range of institutions to verify vaccination certificates that have been issued by a health authority, and are held by the vaccinated individual in question. Issuers are typically institutions such as hospitals, pharmacies, or other entities with the authority to administer vaccinations and accompanying documentation. The holder is the individual who holds the credential with their own vaccination information and can choose to present it for verification to an entity such as an airline or school that might be interested in confirming vaccination status. There have been several proposed paradigms for interactions between these parties. 

Currently widely adopted mechanisms to exchange claims about the end-user require issuers and verifiers of credentials to have a pre-established trust relationship (potentially through an intermediary organization). However, this is unlikely to be effective for vaccine credentials, given that they are expected to be used globally and across contexts in which pre-established trust or relevant intermediaries may not exist. For example, an issuing hospital in Japan cannot be expected to have pre-established trust relationships with all potential verifiers in the US – airports, restaurants, hotels, etc. Thus, the usage of Verifiable Credentials (VCs) has emerged as a promising technology. VCs enable cryptographic verification of the authenticity of the presented proof and can be exchanged relying on ad hoc trust established per transaction, allowing for use across contexts, jurisdictions, and timelines.

For vaccination credentials to be used worldwide, enabling technologies and trust frameworks must have some degree of interoperability. Defining the common set of requirements for the exchange of vaccination credentials has proven to be a challenge, with various stakeholder groups defining their own requirements. In laying out the below principles, we articulate a set of further recommendations towards the architecture and sociotechnical deployment of vaccine credentials.

\section{Access and Equity}

There are a range of potential harms from vaccine credentials that could be precipitated by a rushed implementation, especially for groups that are already marginalized. It is critical that focus stays on ensuring access to vaccines and proof of vaccination to all, so that groups most at risk from COVID for health and economic reasons are not gated on willingness or ability to use a specific credential technology. It is not possible to have an equitable implementation of vaccine credentials without equitable access to vaccines.

A clear plan for educating the public on these credentials, their scope and privacy guarantees, and the basics of their use, should be developed concurrently with the technology, rather than cobbled together post-development. A crucial piece of this could be transparency and notice mechanisms on tools that are already in use. Building functional privacy and data agency into the technology requires not just theoretical privacy guarantees, but must be combined with usability and access to technology for a deployment plan that succeeds in realizing these privacy and data agency benefits for actual users, and so must be developed in communication with the communities and groups we are aiming to include. Deep collaboration with existing civil society organizations and community groups will be needed to understand and mitigate concerns of inequality. As seen with contact tracing technology, sociotechnical systems surrounding deployment build trust with people through offline means, and are central to adoption and widespread use. 

Equity concerns include how user data may be used in a way that may deepen systemic inequalities. For example, contact tracing technology in Singapore originally promised to only use data for public health purposes, but since has been shared with and used by police.\cite{Cox21} In the US, an investigation revealed that the Care19 smartphone app developed for North and South Dakota’s health departments was sharing user information with both Foursquare and Google, despite this violating the app’s stated privacy policy.\cite{Jumbo} One way to mitigate against such breaches of trust is to put safeguards into the design of the technology that protect against misuse and surveillance-creep. 

Looking to existing or upcoming examples of broad-based credentialing efforts will be central to determining processes for public education. The EU’s Digital Green Certificate ("Green Pass"),  for example, would use a QR-code based system to enable anyone vaccinated against Covid, or who has tested negative, or recently recovered from the virus, to travel across all 27 member states. Accessible and secure certificates for all EU citizens is a key element of the proposed regulation,\cite{EUguidance} but there remain implementation concerns. \newline

\emph{Our Principle: Meaningful and user-friendly information must be provided in user-friendly form along with technology deployment, both to promote participation as well as to ensure informed choice. In addition, we must ensure 100\% interoperability with low-tech and paper credential options so that privileged or tech-savvy communities do not receive disproportionate benefit. We acknowledge there will be some convenience gap, and must take care that this does not spill over into structural consequences for individuals who cannot or choose not to participate in this particular technology. We commit to paying special attention to understanding those who may not want to or be able to participate in the vaccine credentialing process.}

\section{Choice and Consent}

There are a range of ethical concerns that may arise from this technology and deployment strategy. One sensitivity involves concern that wide deployment of easy-to-use vaccine credentials would make it possible to mandate vaccines. This is an incredibly politically sensitive issue, with communities across the country particularly concerned about being excluded from basic public spaces due to lack of vaccination – and by extension, lack of vaccine credentials necessary for access and equity.  In the most extreme cases, we’ve seen extremists directly target technologists and organizations.

These issues are likely to be exacerbated if people are not provided with agency, transparency, and education to understand the process, the technology, and the choices available to them. As has been seen repeatedly with poor notice experiences across IT services, lack of explanation hinders trust and adoption. 

By coordinating across Public Health Authorities (PHAs), the standards community, and civil rights and civil society organizations, a communication and adoption plan can be developed to increase trust, understanding of the tech's intended and safe use, and good outcomes. This is particularly important at a time when public distrust of technology and tech companies is rising, and there is already growing sentiment against government-mandated vaccines and COVID monitoring in general.  

Developing robust choices and controls for credential use and providing practical education of the of those controls and choices will increase trust and adoption. Clear communication will be crucial; individuals must be made fully aware of how their vaccine credential data may be used, whether by their own choice or legal requirement. Individuals should be able to access, maintain, and delete or replace the vaccine credential, and do so in a way that is easy to understand. Marginalized populations may not have the resources to request a physical credential if they lose access to their digital credential.  \newline

\emph{Our principle: We believe that individuals have the right to own and manage their personal health data, and further have the right to decide where and how to store this data. Given this, we are building technology that gives people more control and access to their own personal health data. Additionally, we will not build technology intended as a mandatory vehicle for vaccine credentials. We recognize that we must maintain oversight to ensure that businesses and consumers retain the freedom to choose which credential technology, if any, to adopt, and that no data is shared without consent in this process.}

\section{Technology-Enabled, rather than Technology-First}

The number one priority is to build a holistic vaccine verification system that works well for all users of that system – both individual and institutional. The priority is not to build a certain type of technology to accomplish this goal. The biggest challenge to this is that there has been little to no usability testing of the technology, either involving issuers, holders, or verifiers. Major gaps exist in this space:
\begin{enumerate}
\item	Cryptographically verifiable credentials are new to the verifiers, and new to credential holders, who have to manage and present the credentials for verifications. 
\item	Many VC implementations use cryptography and tech are not largely standardized. There are a range of protocols and applications, across JSON-LD, LD-proofs, RDF, BBS+, etc., that are in use and comply with the W3C standards without themselves being fully standardized. In addition, interoperation with already issued paper credentials is still a work in progress. 
\item	VCs used for verifiable vaccination records (or other clinical data) must be capable of conveying the sometimes-nuanced semantics of health records, which can include optional data elements, jurisdiction-specific codes, "modifier elements" and other complexities that verifiers (or intermediaries aiming to simplify the job of verifiers) must be able to process.

\end{enumerate}

There are a slew of startups in this space, and demand from customers and sales is creating uncomfortable pressure on engineers to build and deploy technology in a rushed and haphazard manner in order to be early to market. Technology supposedly built to support COVID response has consistently shown its brittleness when actually deployed,\cite{MIT21} often increasing burden on healthcare workers and public officials, and leading to public distrust. Avoiding this outcome must be one of our top priorities.  

It is likely that a robust, well-communicated, privacy-preserving and interoperable digital solution is the most scalable way for large-scale access to usable credentials. With appropriate safeguards and care for people, this technology can play a part in allowing a return to life and work in the next stage of the COVID-19 pandemic. \newline

\emph{Our principle: Our primary goal is to build something that works for people and that must also follow our principles. Many top-down approaches to COVID technology have failed catastrophically when deployed in the real world, or have failed to achieve adoption as communities organize to fulfill their own needs with stable and trusted lower-tech options (such as spreadsheets and telephone calls). We are committed to working in coordination with the affected communities and involved institutions, moving carefully and thoughtfully when testing and deploying, and addressing the social, economic, and policy realities of COVID vaccinations.}

\section{Privacy and Security}

One of the major strengths of verifiable credentials is their emphasis on privacy; this is one of the clear benefits over paper certificates with no technology component, such as a QR code. The minimal amount of data is used to create the credential, which is generally a subset of the data collected and/or used to administer vaccines. 

Credential holders will need to understand what the privacy and security trade-offs are, via transparency and education, as well as through education from the Public Health Authority (PHA). PHAs can both drive regulatory protections for credential holders and share that information with people, and share information about the benefits and risks of the different credential tools. Design of the technology should include user interfaces that make clear what data is being shared, and easily allow a person to share the minimal amount necessary.

One mechanism for verification is through public-private keys – the verifier will be able to calculate whether an issuer has signed a particular document with their private key, through knowing the issuer’s public key. A fundamental vulnerability is the compromise of a private key. Another potential vulnerability is that if verifiers collude, they can track a user’s movements. There are several ways to address these concerns, such as pairwise identifiers.  
Another common distinction between types of credentials is whether Zero Knowledge Proofs (ZKPs) are integrated, which allow assertions to be verified about a credential without giving all the information in a credential (for example, calculating that someone is over 21 without seeing their age). This is one extreme of a spectrum of technologies to limit how much information is exposed when showing a credential, which range from only being able to show the entire credential, being able to only show specific bits of a credential, and being able to show just a fact that can be derived from the credential. The extent to which this creates or reduces privacy concerns also depends on the use case. However, the ZKPs have not been implemented at scale yet, and the technology is thus largely untested. 

It is not enough just to verify that a credential is legitimately issued, a verifier must also know that the credential belongs to the person in front of them. The simplest way to do this is to have the person’s name, and other datapoints such as an address, included in the credential, allowing the verifier to confirm their identity using traditional methods such as a photo ID. This is yet another area where depending on the use case this may be more or less ideal. For airlines and employers, revealing common attributes of a person's identity isn’t a privacy concern, since the verifier is already aware of these attributes.  Another challenge is that identity verification of vaccine recipients has not been performed robustly during the vaccination process in some countries, including the US and the UK. Individuals are not necessarily being asked to show personal identity documents prior to vaccination, or are being asked to fill in their names on paper-based vaccination credentials themselves.\newline

\emph{Our principle: There are a number of types of digital vaccine credentials and storage methods. For individuals who opt to use systems that store their vaccination data, data collected for vaccine credential purposes should be used only for that purpose, with appropriate safeguards and security. Data should not be shared without permission, and should be deleted as soon as it is no longer needed. The minimal amount of information should be collected and shared as is necessary for the intended use. (See Microsoft’s Privacy Principles for more.)}

\section{Interoperability}
Given the many players in the space, and in the extended universe of healthcare credentialing and digital identity, vaccine credential initiatives are running up against the question of interoperability. 
There are multiple interoperability challenges to be addressed in order for Vaccine Credentials to work, and different groups are focusing on different areas, guided by differences in principles.
\begin{itemize}
\item	\textbf{Interoperability of Adoption.} No matter how perfectly designed a credential system is, it won’t interoperate in practice unless the necessary stakeholders adopt and implement it correctly. One approach to driving adoption would be to legislate mandatory use of a specific verifiable credential, which has been suggested but has significant downsides in preserving choice, consent, and privacy. Instead, regulation of credential technology must be approached with collaboration from those who would be required to implement and use it, to avoid implementation failure and backlash. This will in practice require significant investment in education across relevant institutions, stakeholders, and user communities to ensure an understanding of and access to the technology to facilitate adoption. As of now, there also does not yet exist a plan to interoperate digital verifiable credentials with the paper vaccine proofs and certificates that are currently being handed out to vaccinated individuals. Given that hundreds of millions of individuals have already been vaccinated before  verifiable digital credentials are widely available, this is a significant gap in interoperable adoption. The DIVOC initiative in India does use paper certificates with QR codes as one version of digital credentials,\cite{divoc} but we currently have minimal information on the rollout success of this.  
\item	\textbf{Interoperability of Data.} Ultimately, vaccine credential holders will need to show whether they meet the requirements of being vaccinated. However, the data behind a vaccine credential must include more than a simple “yes” or “no” as the science behind vaccines progresses. Whether a person is considered vaccinated may depend on how long ago they got their vaccine, the efficacy of the vaccine they received on new variants, and other unknowns. The data standards used should align credential issuers and verifiers on the right data in the right format, which will involve close collaboration with medical professionals and public health authorities (at local, national, and international levels) to get health data right, and advocating for adoption of this standard.
\item	\textbf{Interoperability of Trust Framework.} Once the data for a credential is established and securely packaged, it gets issued to the user, who can choose to present it to a verifier. There are a number of ways to accomplish this, each with different privacy and security features depending on the intended scope of use, and this area is being rapidly developed as groups coordinate to develop trust frameworks that include higher degrees of privacy and security applicable to a wide range of sensitive use cases. For example, it may be easier to maintain a list of trusted issuers for those issuers that are large established institutions as well as issuers with resources to spend on getting on the list, and less so for small local clinics and private practices, creating a tradeoff between two outcomes:
\subitem i. Verifying credentials from a plurality of potential issuers may lead to exploitation of the system and false credentials being issued, which in turn could lead to credential fraud. 
\subitem ii. However, being too conservative with the trusted list could lead to smaller or less well funded clinics being left off the list, potentially increasing disparities.

Current initiatives have different views on whether to create fully interoperable credentials and leave it to the verifier to choose to allow certain credentials for their purposes, or whether to curate a list of trusted entities that may issue credentials. Some groups have suggested that coordinating with a federal or state database of vaccinations may be the right answer to this question, but that itself can raise questions of health record centralization, as well as potential concerns for vulnerable groups – not to mention that state and federal governments have a particularly poor track record with COVID healthcare statistics. As the credential ecosystem develops it will likely move toward a trust framework that incorporates more features to safely encompass a larger scope.
\end{itemize}

Regardless of what technical foundation is implemented short-term, it is crucial that standards are developed in partnership with other groups already building and piloting this technology, both for the sake of efficiency and robustness of rollout, as well as to build towards a stronger and more interoperable identity framework in the future. As much as possible, this technology should build on existing standards, given the timeline of standards setting and the urgency of vaccine credentialing efforts. Short-term choices made during this development process may well become part of the foundation for verifiable credential technology moving forward, and may also underpin healthcare technology more broadly. \newline

\emph{Our principle: Health data should be exchanged using a non-proprietary, open, and accessible standard. In the vaccine space, a number of groups are working on developing such standards, across jurisdictions and sectors. Our aim is to move forward in partnership with these organizations, while also interoperating with non-digital credentials, such as the paper certificates already being distributed to current COVID vaccine recipients in the US, and more traditional federated identity solutions. This requires us to actively build trusted relationships and collaborations across involved institutions.}

\section{Continued Transparency, Oversight, and Evolution}
The technology and policy related to vaccine credentials are evolving quickly, with regulations and guidance being proposed and implemented by governments around the world. Small-scale deployments are likely to reveal improvements that are needed to the technology, and as systems are used we will continue to learn about how they function in the real world and affect real people. 

We should also expect to see our scientific understanding evolve as we learn more about covid, vaccines, and the effect of various public health interventions, including where the use of vaccine credentials has positive or negative outcomes. While we are hopeful that there is a place for vaccine credentials in public health policy to make people safer and incentivize staying up to date on vaccinations, this is still very much theoretical. Clumsy implementations of the technology may contribute to backlash or exacerbate existing disparities, leading to greater divisiveness surrounding vaccines. Vaccine credentialing also requires access to vaccines, and we should not contribute to gating services on proving vaccine use without universal vaccine access. 

We also expect a complex interplay between science, technology, and policy. As a society, we have difficult decisions to make regarding what combination of public health interventions and risk we are willing to accept. Some possibilities for return to daily life may include vaccine credentials, but on their own the use of vaccine credentials in one space (such as a workplace or event venue) cannot guarantee low covid risk in that space if there are still high rates of covid in the general population. There remain many unknowns, and we caution against overselling vaccine credentials as a public health solution before research becomes available.

We cannot expect to plan for every problem and design the perfect system from the beginning, but should expect to continue to learn and adapt. Therefore, it is necessary to maintain oversight on all implementations of this nascent technology, listen to community feedback, and be transparent and flexible in our continued learnings as the landscape evolves. We should set up an active process for oversight from relevant stakeholders, particularly affected communities, through civil society partnerships and public transparency on our infrastructure choices and relevant policies. \newline

\emph{Our principle: As with all technology, vaccine credential technology is not neutral. It is often the case that short-term technology built to fulfil a specific, scoped need during a time of urgency can become foundational to the longer-term ecosystem. This requires putting in place firm boundaries surrounding the usage and implementation of vaccine credentials such that privacy is maintained, paired with the necessary oversight to address legal and ethical questions as the ecosystem evolves.}

\printbibliography

\end{document}